# Magnetic anisotropy induced by crystal distortion in Ge$_{1-x}$Mn$_x$Te/PbTe//KCl (001) ferromagnetic semiconductor layers

W. Knoff[a], A. Łusakowski, J.Z. Domagała, R. Minikayev, B. Taliashvili,
E. Łusakowska, A. Pieniążek, A. Szczerbakow, T. Story

*Institute of Physics, Polish Academy of Sciences, al. Lotników 32/46, 02-668 Warsaw, Poland*

[a] corresponding author:  knoff@ifpan.edu.pl

Ferromagnetic resonance (FMR) study of magnetic anisotropy is presented for thin layers of IV-VI diluted magnetic semiconductor Ge$_{1-x}$Mn$_x$Te with $x$=0.14 grown by molecular beam epitaxy (MBE) on KCl (001) substrate with a thin PbTe buffer. Analysis of the angular dependence of the FMR resonant field reveals that an easy magnetization axis is located near to the normal to the layer plane and is controlled by two crystal distortions present in these rhombohedral Ge$_{1-x}$Mn$_x$Te layers: the ferroelectric distortion with the relative shift of cation and anion sub-lattices along the [111] crystal direction and the biaxial in-plane, compressive strain due to thermal mismatch.

## 1. INTRODUCTION

Ge$_{1-x}$Mn$_x$Te is a *p*-type IV-VI diluted magnetic (semimagnetic) semiconductor (DMS) exhibiting carrier-induced ferromagnetism with the Curie temperature $T_C$ depending on both Mn-content $x$ and carrier (hole) concentration $p$ [1,2]. The record values of ferromagnetic transition temperature $T_C$ in $p$-Ge$_{1-x}$Mn$_x$Te of about 150 K for bulk polycrystals [2] and $T_C$=190 K for thin epitaxial layers [3-5] are observed in materials with very high hole concentration $p$≈10$^{21}$ cm$^{-3}$ brought about by electrically active native defects (Ge vacancies). In all known ferromagnetic IV-VI DMS materials with Mn the interspin interactions are driven by the Ruderman-Kittel-Kasuya-Yosida (RKKY) indirect exchange mechanism. The well localized magnetic moments of Mn$^{2+}$ ions (electron configuration 3$d^5$, spin only state with $S$=5/2) are exchange-coupled via conducting holes from Σ- and L- valence bands [6]. The 2+ spin and charge state of Mn ions in IV-VI DMS is observed in both magnetically diluted [7] and concentrated systems [8] as verified by electron paramagnetic resonance (EPR) measurements. As a consequence, Mn ions are in IV-VI semiconductor crystal matrices electrically inactive (replacement of effectively 2+ cations of Pb, Sn or Ge with 2+ Mn ions). As Mn$^{2+}$ ions have no orbital moment ($L$=0) they are only very weakly influenced by ligand crystal electric fields via higher order spin-orbit perturbation effects [9,10]. Therefore, in bulk crystals of cubic (rocksalt) structure these materials show only small magnetocrystalline anisotropy effects. Electronic states derived from magnetic 3$d^5$ (spin up) orbitals of Mn contribute to total density of states of Ge$_{1-x}$Mn$_x$Te deep in the valence band. The photoemission spectroscopy studies showed a peak value about 3.5 eV below the top of the valence band [11,12], i.e. well below the Fermi level positions encountered in actual samples. The spin down states are located well above the bottom of the conduction band and are not occupied.

Bulk GeTe crystallizes in the cubic (rocksalt) structure but well above room temperature (at $T$=670 K) it undergoes a structural transition, in which the high-temperature cubic lattice becomes distorted along the cube's [111] diagonal to form the low-temperature structure having rhombohedral symmetry. One can describe this unit cell by the lattice parameter $a_0$=0.598 nm and the rhombohedral angle $\alpha \approx 88°$ (as compared to $\alpha$=90° for the rocksalt structure) to. The abovementioned transition is accompanied in GeTe by a relative shift $\tau a_0$ of cation and anion sub-lattices along the [111] direction with $\tau \approx 0.03$. This distorted cubic lattice is known to support the ferroelectric state of GeTe [13,14]. Similar structural transition to the ferroelectric state is observed in $Ge_{1-x}Mn_xTe$ as well, but the incorporation of Mn ions stabilizes the cubic phase and the rhombohedral distortion takes place at the temperatures lowered down to room temperature in bulk polycrystals with $x$=0.18 [2] and also in thin layers with $x$=0.2-0.3 grown on $BaF_2$ substrate [3-5, 15-18]. The co-existence of the ferromagnetic and the ferroelectric order in single-crystalline phase of $Ge_{1-x}Mn_xTe$ monocrystals gives opportunity to study their unique multiferroic properties.

Magnetization, anomalous Hall effect and ferromagnetic resonance (FMR) measurements of (0.1-1) micron-thick $Ge_{1-x}Mn_xTe$ layers deposited on $BaF_2$ (111) substrates [3-5,15-18] with Mn concentration $x$<0.2 (i.e. possessing the rhombohedral structure) showed magnetization easy axis oriented along the growth direction (perpendicularly to the layer plane). In a sharp contrast, the layers with higher Mn content (having the rocksalt structure) showed regular easy-plane type magnetic anisotropy. These experimental findings strongly suggested that the ferroelectric distortion along the [111] growth direction is responsible for the perpendicular magnetic anisotropy in $Ge_{1-x}Mn_xTe//BaF_2$ (111). The two key magnetic anisotropy energy contributions: the uniaxial term due to ferroelectric distortion and the cubic magnetocrystalline anisotropy term were used in recent phenomenological quantitative analyses of the temperature and composition dependence of the FMR resonant field [16,18]. It was also concluded that magnetic shape anisotropy originating from dipolar interactions contributes (with the opposite sign) to the uniaxial term.

These experimental observations and conclusions were supported by the very recent density functional theory (DFT) analysis of magnetic anisotropy mechanisms in $Ge_{1-x}Mn_xTe$ [19,20]. The perpendicular magnetic anisotropy was understood in the mechanism involving the hybridization of magnetic $3d^5$ electronic orbitals of $Mn^{2+}$ ions and valence band states in GeTe resulting in induced spin polarization of neighboring nonmagnetic ions of Te and Ge. The theoretical calculations identified two most important physical aspects of the magnetic anisotropy mechanism: the spin-orbital interactions in the valence band states and the distortion of the cubic crystal lattice [19,20].

In this work, we employ the FMR method to study the magnetic anisotropy in $Ge_{1-x}Mn_xTe$ ($x$=0.14) layers deposited on KCl (001) substrate. For Mn content of 14 at. %, a layer is expected to grow in the rhombohedral structure and to exhibit the ferroelectric [111] distortion discussed above. Additionally, due to very large mismatch between the thermal expansion coefficients of the KCl substrate and IV-VI semiconductors, the $Ge_{1-x}Mn_xTe$ layer will at low temperatures be under strong biaxial $x$[100]-$y$[010] compressive strain with the corresponding tensile strain along the growth direction $z$[001] [21-24]. This biaxial distortion is, for example, known to result in a substantial decrease of the band gap (by 70 meV) of IV-VI semiconductor layers on KCl substrate [21-23] and in an increase of the strength of interspin exchange interactions and the ferromagnetic Curie temperature in closely related IV-VI EuS-PbS//KCl (001) multilayers [24]. We analyzed the angular dependence of the FMR resonant field in various crystal planes. By determining the symmetry of anisotropic contribution to total magnetic energy of the layer we identified two uniaxial magnetic energy terms with easy magnetization axes along [111] and [001] directions We found the magnetization easy axis controlled by these two distortions with the angular dependence of magnetic anisotropy energy possessing a broad minimum near the normal to the layer plane. The experimental results for $Ge_{1-x}Mn_xTe$ layers grown on KCl (001) substrates are compared with the studies of the layers deposited on $BaF_2$ (111) substrates, in which all dominant distortion sources (ferroelectric and thermal mismatch) have the same main axes and the magnetic easy axis is normal to the layer plane. The aim of our work is to verify experimentally the proposed microscopic model of magnetic anisotropy in $Ge_{1-x}Mn_xTe$ and to demonstrate, how to control the magnetization easy axis by engineering of the crystal lattice distortions.

## 2. LAYER GROWTH AND CHARACTERIZATION

$Ge_{1-x}Mn_xTe$ layers having thickness of about one micron were grown by molecular beam epitaxy (MBE) technique on freshly cleaved insulating and nonmagnetic KCl (001) substrates at the temperature $T=350$ °C using GeTe, Mn, and $Te_2$ molecular sources. In order to obtain high crystalline quality epitaxial heterostructures, we carried out a few tenths of test growth processes depositing the layer either directly on KCl (001) surface or employing a proper IV-VI buffer layer of GeTe or PbTe (buffer thickness in the range 20 – 200 nm). The layer growth was monitored in situ by the reflection high-energy electron diffraction (RHEED) technique. In the optimized regime, the RHEED revealed a clear, streaky pattern characteristic for the two-dimensional mode of growth in the case of the PbTe buffer and a sharp spotty pattern indicating a three-dimensional island–type growth mode of the $Ge_{1-x}Mn_xTe$ layer. The Mn content in the samples studied ($x=0.12$ - $0.15$) and their chemical homogeneity were examined in a SEM microscope by energy-dispersive X-ray (EDX) fluorescence analysis. The atomic force microscopy (AFM) analysis of the surface morphology of the top $Ge_{1-x}Mn_xTe$ layer revealed the mean roughness parameter of 3 nm.

Due to fairly large lattice parameter mismatch $\Delta a/a \approx 2.6$ % between the KCl substrate ($a_0=0.629$ nm) and the PbTe buffer ($a_0=0.646$ nm), as well as the 7.4 % mismatch between the buffer and the GeTe layer ($a_0=0.598$ nm), the critical thickness for the growth of pseudomorphically strained GeTe/PbTe//KCl heterostructure remains below 10 nm [21,22]. Therefore, the micron-thick GeTe and $Ge_{1-x}Mn_xTe$ layers on the PbTe//KCl (001) substrates are expected to be fully relaxed by formation of interfacial defects. However – due to very large mismatch between thermal expansion coefficients $\alpha_T$ of the substrate ($\alpha_T = 3.9*10^{-5}$ $K^{-1}$ for KCl at room temperature) and IV-VI semiconductors (typical value $\alpha_T = 2*10^{-5}$ $K^{-1}$) – a $Ge_{1-x}Mn_xTe$ layer is at low temperatures expected to be under strong biaxial x-y compressive strain of the order of $\varepsilon \approx 1$ % [21,22]. The corresponding tensile strain along the growth direction $z$ is determined by the Poisson coefficient $\Delta a_z/\Delta a_{x,y} \approx -(0.15 - 0.45)$ in IV-VI semiconductors [21-24]. This is the main source of strain expected to further lower the crystallographic symmetry the of GeTe/PbTe//KCl (001) layers. The lattice distortion due to this strain mechanism can be evaluated as thermal expansion induced relative change of lattice parameters integrated between the growth temperature and the experimental one: $\Delta a_{x,y}/a = \int \Delta \alpha_T(T) dT$. This mechanism is particularly effective at low temperatures, when the full strain can build up because of negligible thermodynamic probability of strain release by formation of crystal defects [21,22].

Structural properties of $Ge_{1-x}Mn_xTe$/PbTe//KCl heterostructures were examined by the X-ray diffraction (XRD) method at room temperature demonstrating good quality monocrystalline layers of both PbTe and $Ge_{1-x}Mn_xTe$. The layers grew epitaxially and propagate the (001) crystal orientation of the substrate. Fig. 1 presents the standard $2\theta$ -$\theta$ XRD diffractogram, in a semi-logarithmic intensity scale, obtained using powder diffractometer equipped with Johanson monochromator and stripe detector.

No other crystal phases or crystal orientations were detected for the layers studied in this work. The sequence of three peaks corresponds to three interplane distances (along the [001] growth direction) as expected from the lattice parameters of the materials: $a_{PbTe} > a_{KCl} > a_{GeMnTe}$. In order to verify the expected strain-induced distortion, we employed high resolution X-ray diffractometer (HR-XRD) in a geometry sensitive for either symmetric or asymmetric Bragg reflections. The lattice parameters along selected crystallographic axes were determined. Our measurements provide the clear evidence for deformation of the unit cell of $Ge_{0.86}Mn_{0.14}Te$ with the in-plane ($a_{x,y}$=0.5888 nm) and the out-of-plane ($a_z$=0.5953 nm) lattice parameters differing considerably. These XRD experimental findings agree well with the abovementioned growth scenario for $Ge_{1-x}Mn_xTe$ layer on KCl substrates and provide clear evidence for the presence of the in-plane biaxial compressive strain and the out-of-plane Poisson tensile strain. However, for these distorted $Ge_{1-x}Mn_xTe$ unit cells we could not provide a direct XRD evidence for the cation-anion sub-lattice shift along the [111] direction in $Ge_{0.86}Mn_{0.14}Te$, which is expected based on the phase diagram of both bulk crystals and thin layers on $BaF_2$ (111) substrates.

The full angular dependence of magnetic anisotropy in (Ge,Mn)Te layers was studied by the FMR technique with a Bruker X-band spectrometer operating at the microwave frequency of $v$=9.4 GHz. This device is equipped with a continuous flow helium cryostat covering the temperature range $T$=3–300 K and a goniometer for the control of the polar angle of applied magnetic field $\theta_H$ with the accuracy of about 1°. The ferromagnetic resonance is observed as a characteristic temperature and angular dependence of the resonant field and resonance intensity as shown in Fig. 2. Below the ferromagnetic transition temperature of about $T_C$=50 K a strong decrease of the resonant field is observed from the value corresponding to the g-factor of 2.0 expected for $Mn^{2+}$ ions in IV-VI DMS semiconductors [7,8]. As the FMR lines observed in $Ge_{1-x}Mn_xTe$ are fairly broad having the peak-to-peak width $\Delta H_{pp} \approx 1$ kOe, we estimate the accuracy of determination of the resonant field to be of ±50 Oe.

## 3. EXPERIMANTAL RESULTS AND DISCUSSION

The angular dependence of the FMR resonant field observed in $Ge_{0.86}Mn_{0.14}Te/PbTe//KCl$ layers is presented in Fig. 3. During the measurements the samples were rotated by 360° (with the step of 10°) from [001] direction ($\theta_H$=0°, normal to the layer) either to [1-10] direction or to [010] direction in the (001) layer plane (see schematics in Fig. 3). The angle $\theta_H$=90° corresponds to the magnetic field applied in the plane of the layer. Maxima and minima of the FMR resonant field clearly identify the hard and easy axes of magnetization. In contrast to the expectations based on shape anisotropy mechanism, the easy magnetization axis in $Ge_{0.86}Mn_{0.14}Te/PbTe//KCl$ is not located in the layer plane, but it is close to the layer normal with a broad magnetic anisotropy energy minimum around $\theta_H$=0°. This effect is quite strong (maximum difference of resonant fields $\Delta H_R$ = 1.6 kOe) and is observed for both (100) and (-110) magnetic field rotation planes (see Fig. 3a). Surprisingly strong anisotropy of 180° period is also observed for magnetic field rotation in the plane of the layer ($\Delta H_R$ = 0.4 kOe). We will show below that both effects stem from the same origin. In Fig. 4 we compare the magnetic anisotropy observed for similar $Ge_{1-x}Mn_xTe$ layers deposited by MBE on $BaF_2$ (111) (our recent work [18]) and KCl (001) substrates (this work). The perpendicular magnetic anisotropy is observed for both layers with a marked difference in maximum (hard axis) location: strictly at $\theta_H$ = 90° for the layer on $BaF_2$ but at a bigger angle $\theta_H$ > 90° for the layer on KCl. The solid lines in Fig. 4 describing the experimental data were calculated based on a magnetic anisotropy model presented below.

The experimental data were analyzed assuming the following form of the free energy of magnetic anisotropy for (001) oriented layers:

$$F = K_{4c}\left(\alpha_x^4 + \alpha_y^4 + \alpha_z^4\right) + D_1\alpha_z^2 + \frac{1}{3}D_2\left(\alpha_x + \alpha_y + \alpha_z\right)^2 - \vec{H}\vec{M} \qquad (1)$$

where $\alpha_x$, $\alpha_y$ and $\alpha_z$ are the directional cosines in coordinate system with the main axes along <001> crystallographic directions. The standard Zeeman term $\vec{H}\vec{M}$ is supplemented with the terms proportional to $K_{4c}$, $D_1$ and $D_2$ describing cubic, tetragonal, and trigonal anisotropies, respectively. The presence of tetragonal anisotropy is related to the thermal strain induced difference between in-plane and out-of-plane lattice parameters in $Ge_{1-x}Mn_xTe/PbTe//KCl$. The $D_1$ term effectively contains also the positive contribution of shape anisotropy that possesses identical symmetry. The trigonal term $D_2$ takes into account the ferroelectric distortion of $Ge_{1-x}Mn_xTe$ ($x<0.2$) crystals along [111] crystallographic direction. The FMR resonant field was calculated from the standard free energy derivative equation (2):

$$\left(\frac{\hbar\omega}{g\mu_B}\right)^2 = \frac{1}{M^2\sin^2\theta}\left(F_{\theta\theta}F_{\varphi\varphi} - F_{\theta\varphi}^2\right) \qquad (2)$$

Here $\omega$ is the microwave frequency, $g$ is the g-factor of Mn ions, $M$ is magnetization, $h$ is the Planck constant, and $\mu_B$ is the Bohr magneton.

Fig. 5 illustrates the influence of various terms in the free energy (1) on the angular dependence of the FMR resonant field. It was assumed in the calculations that the g-factor of Mn ions is $g = 2.73$ at helium temperatures. This value provides the best overall fit to our data on the anisotropy of the FMR resonant field. The surprisingly large value was already found in previous FMR studies of $Ge_{1-x}Mn_xTe$ on $BaF_2$ substrate [16,18] but remains not fully understood. Several proposals were considered, like exchange-coupling between carrier-rich ferromagnetic and carrier-poor antiferromagnetic regions in $Ge_{1-x}Mn_xTe$ layers (electronically nonhomogeneous at nanoscale) or the coupling to spin-orbital valence band states. We note that the $g = 2.0$ is recovered at high temperatures (see Fig. 2). The continuous, broken and dotted lines correspond to the external magnetic field in the (100), (-110) and (001) crystallographic planes, respectively. For the (100) and (-110) planes the angle is measured from the [001] crystallographic direction, while for the (001) plane (the layer plane) it is measured from the [100] direction. In all panels, the coefficient describing tetragonal deformation $D_1 = -550$ Oe. This term contains also the $2\pi M$ shape anisotropy term of about +250 Oe (expected based on magnetometry data). Therefore, the tetragonal deformation induced a term, which is of about -800 Oe and constitutes the dominant magnetic anisotropy contribution in $Ge_{1-x}Mn_xTe/PbTe//KCl$.

In Fig. 5a, we consider the reference case with $D_1 = -550$ Oe, but the cubic and the ferroelectric terms equal zero, $D_2=K_{4c}=0$. One can notice symmetry of the curves with respect to $\theta_H = 90°$, i.e. for magnetic field along the [010] direction. The continuous and broken lines are the same and no in-plane anisotropy is observed. The situation changes when we add cubic anisotropy term in Fig. 5b, $K_{4c}=50$ Oe. Although the curves remain symmetric with respect to the angle $\theta_H = 90°$, there is a difference between (100) and (-110) planes (crystal field effect). Moreover, we observe the weak in-plane angle dependence of the resonant field with the period of 90°. The symmetry with respect to $\theta_H = 90°$ disappears in Fig. 5c and Fig. 5d, where we allow for the trigonal deformation along [111] direction, $D_2 = -300$ Oe. In Fig. 5c, the crystal field effect is zero ($K_{4c}=0$), whereas in Fig. 5d $K_{4c}=50$ Oe and all three anisotropy terms are present. The in-plane anisotropy is now much stronger and exhibits the period of 180°. It clearly suggests that this experimentally observed effect (see Fig. 3b) originates not from the usual cubic crystal field anisotropy but stems from the ferroelectric distortion being in (001) oriented layers inclined from the normal to the layer. The in-plane anisotropy is practically absent in $Ge_{1-x}Mn_xTe$ layers grown on $BaF_2$ (111), when the ferroelectric distortion acts along the normal to the layer [18]. We note that – although the calculated strength and period of the in-plane anisotropy agree well with our experimental

observations – there is a 45° phase shift between our experimental data and calculations. This effect requires further studies.

The signs of the coefficients $D_1$, $D_2$ and $K_{4c}$ are important in analyzing the experimental data. The coefficient $D_1$ must be negative for the [001] crystallographic direction to be the easy axis of magnetization. The term $D_2$ also must be negative to reproduce the experimental observation that the anisotropy in the (-110) plane (broken curve) is smaller, than the anisotropy in the (100) plane (continuous curve) as well as for the maxima of both curves to appear at the polar angle $\theta_H > 90°$, precisely like in the experiment.

In the model discussed above, we successfully explained the key experimental features of magnetic anisotropy of $Ge_{1-x}Mn_xTe$ layers on KCl (001) substrates, in particular the nearly perpendicular to the layer location of magnetization easy axis and the origin of strong in-plane anisotropy. We would like to note, however, that throughout the calculations we adopted a specific [111] direction of the ferroelectric distortion, neglecting the other <111> directions that are symmetry-equivalent in cubic and tetragonal crystals. In doing so, one neglects the possible ferroelectric domain structure with ferroelectric distortion along various <111> directions. Although the crystal symmetry is even lower in our $Ge_{1-x}Mn_xTe$ layers on KCl (001) substrates, one may expect two possible directions of ferroelectric distortion present in the crystal rotation planes chosen in our experiments. This problem was recently discussed in ref. [16] for $Ge_{1-x}Mn_xTe$ layers on $BaF_2$ (111) substrate taking into account the magneto-electric coupling of electrical polarization (deformation) and magnetization vectors and magnetic field dependent contributions of various domains. Discussion of these effects is beyond the scope of this work.

## 4. CONCLUSIONS

Applying the ferromagnetic resonance (FMR) method, we studied magnetic anisotropy in micron-thick monocrystalline layers of ferromagnetic $Ge_{0.86}Mn_{0.14}Te$ IV-VI semiconductor deposited by molecular beam epitaxy on KCl (001) substrate with a 20 nm-thick PbTe buffer layer. The analysis of the angular dependence of the FMR resonant field was carried out taking into account the standard magnetic energy contributions of Zeeman, cubic magnetocrystalline and shape anisotropy terms supplemented with two uniaxial terms due to the ferroelectric rhombohedral distortion along [111] direction and the thermal strain induced, tetragonal biaxial distortion along [001] crystal axis. The easy magnetization axis was found to be near to the normal to the layer plane as set by dominant magnetic anisotropy terms due to the tetragonal and the ferroelectric uniaxial distortions. The ferroelectric distortion has been recently experimentally and theoretically identified as the origin of perpendicular magnetic anisotropy in rhombohedral $Ge_{1-x}Mn_xTe$ layers ($x<0.2$) deposited on $BaF_2$ substrates with [111] growth direction. In contrast, we showed that in $Ge_{1-x}Mn_xTe$/PbTe//KCl (001) layers (apart from the ferroelectric distortion) also the thermal mismatch induced in-plane compressive distortion accompanied by tensile distortion along the [001] growth direction contributes to magnetic anisotropy, thus confirming the recently proposed microscopic mechanism. It shows that various lattice deformations can be used to engineer the required magnetic anisotropy and the specific location of easy magnetization axis in ferromagnetic $Ge_{1-x}Mn_xTe$ semiconductor heterostructures.

**ACKNOWLEDGMENTS**

This work was supported by the research project UMO 2011/01/B/ST3/02486 of Polish National Science Centre (NCN).

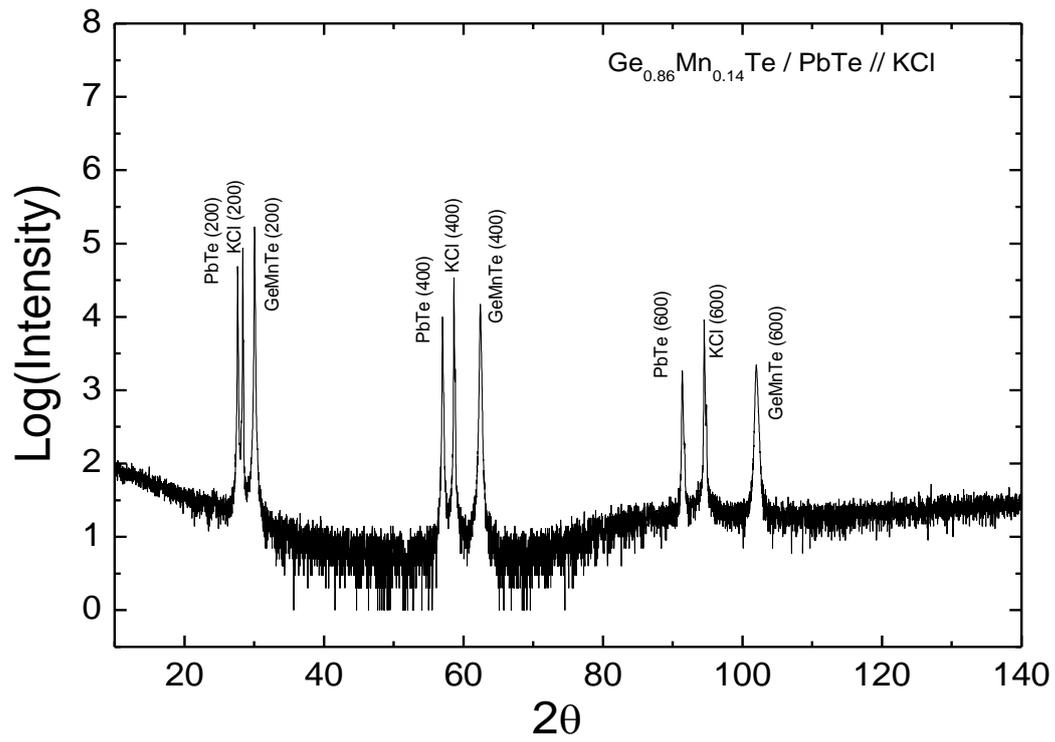

FIG. 1. Standard X-ray diffraction $2\theta$-$\theta$ scan at room temperature for $Ge_{0.86}Mn_{0.14}Te/PbTe$ layer grown on KCl (001) substrate.

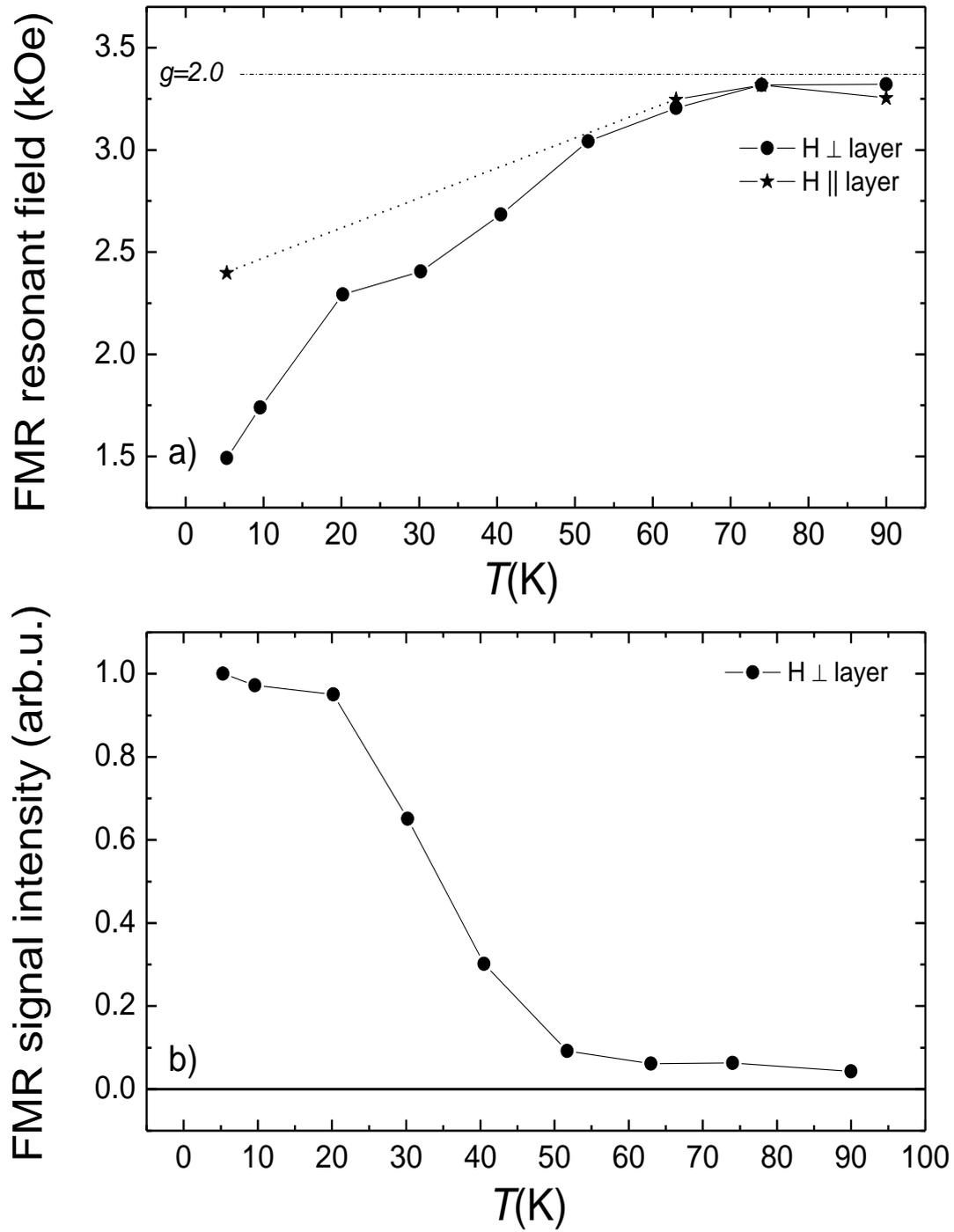

FIG. 2. Temperature dependence of the FMR resonant field (a) and integrated intensity (b) obtained for external magnetic field applied perpendicular (circles) or parallel (stars) to the $Ge_{0.86}Mn_{0.14}Te/PbTe//KCl$ (001) layer.

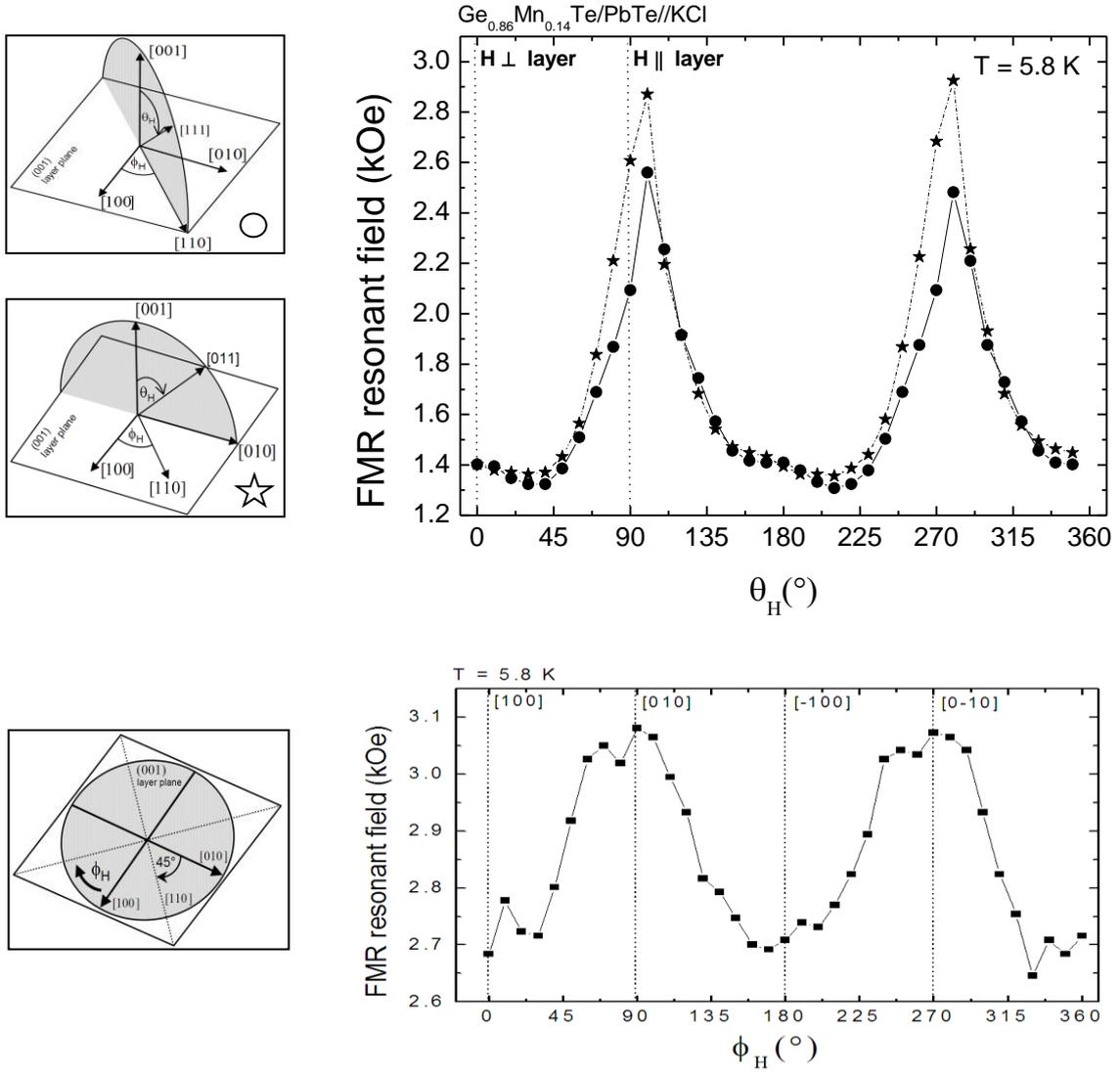

FIG. 3. The angular dependence of the FMR resonant field in $Ge_{0.86}Mn_{0.14}Te/PbTe//KCl$ (001) layer for three rotation planes of the layer with respect to the external magnetic field as shown in the figure. Figure 3a shows data obtained for two magnetic field rotation planes normal to the layer: the (1-10) plane (circles) and the (100) plane (stars). Figure 3b presents results for the sample rotated in the (001) plane. The lines are the eye-guides only.

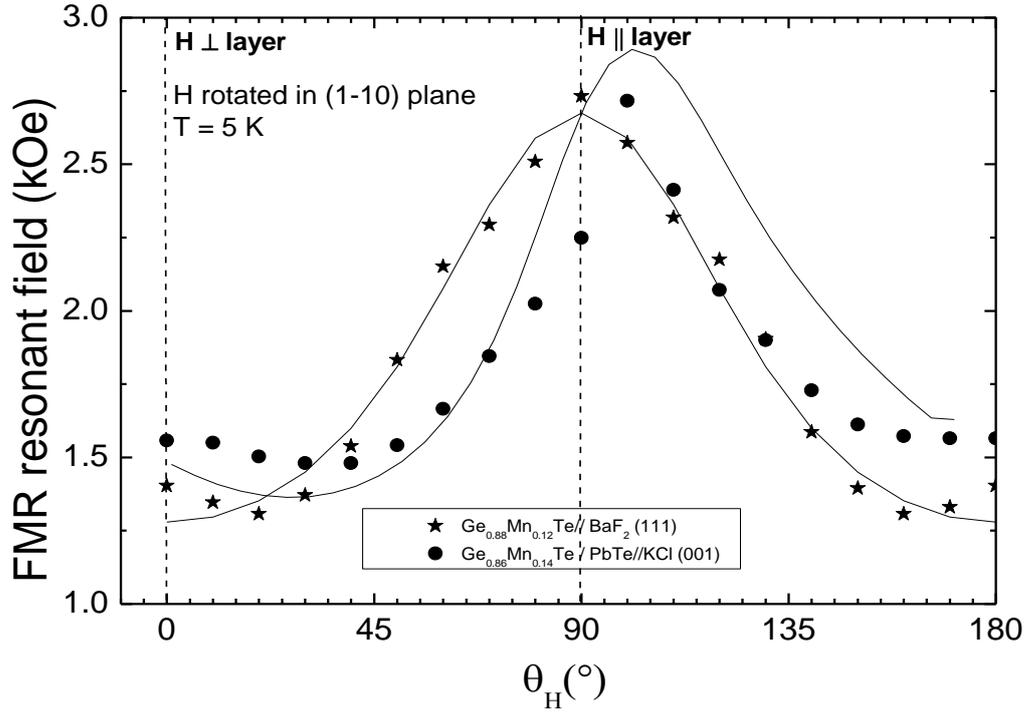

FIG. 4. Comparison of the angular dependences of the FMR resonant field for $Ge_{0.88}Mn_{0.12}Te//BaF_2$ (111) layers (stars) and $Ge_{0.86}Mn_{0.14}Te/PbTe//KCl$ (001) layers (circles). In both cases sample's rotation takes place in (1-10) plane. The in-plane $\theta_H=90°$ corresponds to [110] direction for KCl (001) substrate but to [11-2] direction for $BaF_2$ (111) substrate. Solid lines show theoretical calculations based on the analysis of magnetic anisotropy energy with distortion-induced uniaxial terms discussed in the text.

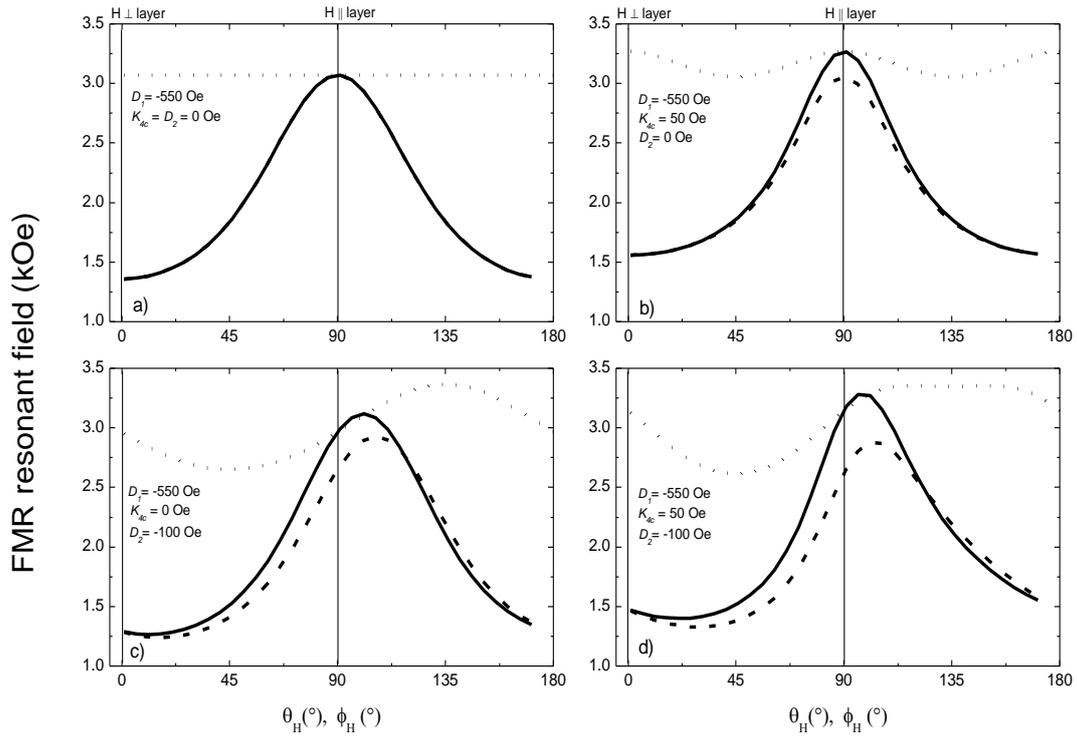

FIG. 5. Theoretical calculations of the angular dependence of the FMR resonant field based on free energy analysis described by equation (1). Fig 5a – 5d shows various scenarios of magnetic anisotropy in $Ge_{0.86}Mn_{0.14}Te/PbTe//KCl$ taking into account the cubic ($K_{4c}$), tetragonal ($D_1$) and trigonal ($D_2$) contributions. In all calculations effective g-factor at $T= 5$ K was $g=2.73$. The continuous and broken lines correspond to the rotation in the (100) and (-110) crystallographic planes (polar angle $\theta_H$), respectively. The dotted lines correspond to the rotation in the (001) layer plane as determined by azimuthal angle $\phi_H$.